\documentstyle[preprint,psfig,tighten,floats,aps]{revtex}
\def\ut#1{\rlap{\lower1ex\hbox{$\sim$}}{#1}}

\def\pb#1{\rlap{\lower1.5ex\hbox{$\longleftarrow$}}{#1}}
\def\dpb#1{\rlap{\lower1.5ex\hbox{$\Longleftarrow$}}{#1}}
\def\spb#1{\rlap{\lower1.0ex\hbox{$\leftarrow$}}{#1}}
\def\sdpb#1{\rlap{\lower1.0ex\hbox{$\Leftarrow$}}{#1}}

\def\d{{\rm d}}

\def\ba{\begin{eqnarray}}
\def\ea{\end{eqnarray}}
\def\be{\begin{equation}}
\def\ee{\end{equation}}

\def\={\mathrel{\widehat\mathalpha{=}}}
\def\puto#1{\rlap{\raise.5ex\hbox{\char'27}}{#1}}

\preprint{\vbox{\baselineskip=12pt
\rightline{ICN-UNAM-01-13}}}

\begin{document}
\draft
\title{A Mass Formula for EYM Solitons}
\author {Alejandro Corichi${}^{1}$\thanks{E-mail address: 
corichi@nuclecu.unam.mx}, Ulises Nucamendi${}^{2}$ and
Daniel Sudarsky${}^{1}$
}

\address{1. Instituto de Ciencias Nucleares\\
Universidad Nacional Aut\'onoma de M\'exico\\
A. Postal 70-543, M\'exico D.F. 04510, M\'exico.}

\address{2. Center for Theoretical Physics\\
University of Sussex\\
Brighton BN1 9QJ, Great Britain.
}

\maketitle

\begin{abstract}
The Isolated Horizon formalism, together with a simple
phenomenological model for colored black holes was
recently used to predict a
formula for the ADM mass of the solitons of the EYM system
in terms of horizon properties
of black holes {\it for all} values of the horizon area. In this note,
this formula is
tested numerically  --up to a large value of the area-- for spherically
symmetric solutions and shown to yield the known masses of the solitons.
\end{abstract}

%\section{Introduction}
%\label{sec1}

Historically, the use of simple models has played an  important role to
gain intuition about the behavior of complex gravitational systems.
For example, the intuitive notion of ``what can be radiated {\it will}
be radiated'' after a gravitational collapse gave way to the formulation
of the ``no-hair conjecture'' for stationary solutions
\cite{Chrusciel:1994sn}. Similarly,
the fact that there exist no solitonic (i.e. stationary, regular) solutions
to the Yang-Mills equations on Minkowski space-time was in part responsible for
the general belief that solitons were also absent in the presence of
gravity. However, as these examples illustrate, the intuitive models that
one constructs are of a limited validity once one considers more general
situations. For instance, it is now known that the Einstein-Yang-Mills
system violates both of the early ``conclusions'', that is, there exist
both solitons \cite{Bartnik:1988am,Smoller:1993pe} 
and colored black holes \cite{Bizon:1990sr,Smoller:1993bs}
in that theory.
More recently, the use of dimensional
analysis has helped to develop intuition about the existence of static
(regular and BH type) solutions in theories with different matter
content \cite{Bizon:1994dh}.
The basic idea is that whenever a dimension-full constant can be constructed
from the fundamental coupling constants of the theory, then there exist
non-trivial static solutions. For example, in the EYM system the gravitational
constant $G$ and the gauge coupling constant $g$ give raise to a mass
parameter (or a length scale) that governs the scale of the mentioned
solutions. One of the facts that can be heuristically explained from
this viewpoint is the emergence of new phenomena once more fields (and thus
more coupling constants) are introduced. For instance, in the passage from
EYM to EYM-Higgs, a new mass parameter can be constructed from the
expectation value $\eta$ of the Higgs field. This indicates, for
instance, that the hair of the hairy black holes is always short and that
those BH have an upper bound on their area \cite{Nunez:1996xv}.

During the past year, the introduction of the Isolated Horizon (IH)
formalism \cite{Ashtekar:2000yj,Ashtekar:2000hw} has allowed to
gain insight into the static sector of theories admitting ``hair''
\cite{Corichi:2000dm}.  First,
it was found that the Horizon Mass of the black hole, constructed
purely of quasi-local quantities, is related in a simple way
 to the ADM masses of
both colored black holes and solitons of the theory
\cite{Corichi:2000dm}. Second,
a simple model of colored black holes as bound states of regular black
holes and
solitons emerged and  provided heuristic explanations for the behavior of
horizon quantities of colored black holes \cite{Ashtekar:2001nx}.
Finally, the combination of
the Mass formula, together with the fact that in EYMH theory, different
``branches'' of static solutions merge, has lead to a formula for the
difference of soliton masses in terms of black hole quantities
\cite{Kleihaus:2000kv,Ashtekar:2001nx}.
In this article, we explore further the consequences of the IH formalism
in the static sector of the theory. In particular,
we explore  a formula for the ADM mass of the $n$ Bartnik-McKinnon soliton
\cite{Bartnik:1988am} in terms of horizon quantities of the
related black holes,
\be
M_{\rm sol}^{(n)}=\frac{1}{2G}\int_0^\infty (1-\beta_{(n)}(r)) \d r
\label{Mass}
\ee
where $\beta_{(n)}(r)=2r\kappa_{(n)}(r)$ and $\kappa_{(n)}(r)$ is the
surface gravity of the $n$ colored black hole of area $a_\Delta=4\pi r^2$.
This  formula is striking because it allows to compute masses of {\it solitons}
from surface gravity of {\it black holes}. It should be noted that even when
this formula refers to the EYM system, it was predicted from considerations
in the EYMH system \cite{Ashtekar:2001nx}.

Two basic ingredients
are needed for the validity of this formula: i) That {\it  the
static sector} of the EYMH system, in the limit when the 
vacuum expectation value $\eta$ approaches zero  ($\eta\mapsto 0$) 
reduces to the static sector of the EYM
system. Thus, the solutions of the EYM system would correspond to
 limiting solutions of EYMH equations; 
ii) The surface gravity
of colored black holes should have certain limiting behavior
as the horizon area grows (in
order to have convergence of the integral in (\ref{Mass})).
In this note  we shall prove i) and,
by numerical integration, we  show evidence for  the
mass formula (\ref{Mass}), and thus
indirectly of the limiting behavior of surface gravity.
Our numerical integration is restricted to the spherically symmetric sector
and
for the $n=1,2$ colored black holes, but one expects the main result to be
valid in general, non-spherical solutions and for all values of $n$.

%\section{Preliminaries}
%\label{sec2}

Let us begin by recalling the
{\it Isolated Horizons} (IH) formalism, whose applications range from the
extraction of physical quantities in numerical relativity to
quantum entropy calculations\cite{Ashtekar:2000sz}.
The basic idea is to consider space-times with an interior boundary
(to represent the horizon), satisfying quasi-local boundary conditions
ensuring that the horizon remains `isolated'. Although the boundary
conditions are motivated by geometric considerations, they lead to a well
defined action principle and Hamiltonian framework.
Furthermore, the boundary
conditions imply that certain `quasi-local charges', defined at the horizon,
remain constant `in time', and can thus be regarded as
the analogous of the global charges defined at infinity in the
asymptotically flat context. The isolated horizon Hamiltonian framework 
enables one to define the {\it Horizon Mass} 
$M_\Delta$, as function of the `horizon charges'.

In the Einstein-Maxwell and Einstein-Maxwell-Dilaton
systems considered originally \cite{Ashtekar:2000yj},
the horizon mass satisfies a
Smarr-type formula and a generalized first law in terms of
quantities defined exclusively at the horizon (i.e. without any reference to
infinity).
The introduction of non-linear matter fields like the Yang-Mills field
brings unexpected subtleties to the formalism\cite{Corichi:2000dm}.
However, one can still define a Horizon Mass, and
furthermore, this Horizon Mass satisfies a first law.

An isolated horizon is a non-expanding null surface generated by a
(null) vector field $l^a$. The IH boundary conditions imply that
the acceleration $\kappa$ of $l^a$ ($l^a\nabla_al^b=\kappa l^b$)
is constant on the horizon $\Delta$. However, the precise value it takes
on each point of phase space  is not determined a priori. On the
other hand, it is known that for each vector field $t^a_o$ on space-time,
the induced vector field $X_{t_{o}}$ on phase space is Hamiltonian
if and only if there exists a function $E_{t_{o}}$ such that
$\delta E_{t_{o}}=\Omega (\delta,X_{t_{o}})$, {\it for any
vector field $\delta$ on phase space}. This condition can be re-written
 as\cite{Ashtekar:2000hw}, $
\delta E_{t_{o}}=\frac{\kappa_{t_{o}}}{8\pi G}\,\delta a_{\Delta}
+ {\rm work\;\; terms}$.
Thus, the first law arises as a necessary and sufficient condition for
the consistency of the Hamiltonian formulation;  the allowed vector
fields $t^a$ will be those for which the first law holds.
Note that there
are as many `first laws' as allowed vector fields $l^a\=t^a$ on the horizon.
However, one would like to have a {\it physical first law}, where the
Hamiltonian $E_{t_{o}}$ be identified with the `physical mass' $M_{\Delta}$
of the horizon. This amounts to finding the `right $\kappa$'.
This `normalization problem' can be easily overcome in
the EM system \cite{Ashtekar:2000yj}. In this case, one chooses the function
$\kappa=\kappa(a_\Delta, Q_\Delta)$ as the corresponding
function for the {\it static}
solution with charges $(a_\Delta, Q_\Delta)$.
However, for the EYM system, this procedure is not as straightforward.
A consistent viewpoint is to  abandon  the notion of
a globally defined  horizon mass on Phase Space,
and to define, for each value of $n=n_o$,
a canonical normalization  $t^a_{n_o}$ that yields the Horizon Mass
$M^{(n_o)}_{\Delta}$
for the $n_o$ branch \cite{Ashtekar:2000hw}. The horizon mass takes the form,
\be
M^{(n_o)}_{\Delta}(R_\Delta)=\frac{1}{2 G}\int_0^{R_\Delta}\beta_{(n_o)}(r)
\, \d r\, ,
\ee
along the $n_o$ branch with $R_\Delta$ the horizon radius.

Finally, for any $n$ one can relate the horizon mass $M^{(n)}_{\Delta}$
to the ADM mass of static black holes.  Recall first that general
Hamiltonian considerations imply that the total Hamiltonian,
consisting of a term at infinity, the ADM mass, and a term at the horizon,
the Horizon Mass, is constant on every connected component of static
solutions (provided the evolution vector field $t^a$ agrees with the
static Killing field everywhere on this connected component)
\cite{Ashtekar:2000yj,Ashtekar:2000hw}.
In the Einstein-Yang-Mills case,
since the Hamiltonian is constant on any $n$-branch, we can evaluate
it at the solution with zero horizon area.  This is just the soliton,
for which the horizon area $a_\Delta$, and the horizon mass $M_{\Delta}$
vanish.  Hence we have that $H^{(t_0,n)} = M_{\rm sol}^{(n)}$.
Thus, we conclude:
\be \label{ymmass}
M^{(n)}_{\rm sol} = M^{(n)}_{\rm ADM} - M^{(n)}_{\Delta} \ee
on the entire $n$th branch \cite{Corichi:2000dm,Ashtekar:2000hw}.
Thus, the ADM mass
contains two contributions, one attributed to the black hole horizon
and the other to the outside `hair', captured by the `solitonic
residue'. The formula (\ref{ymmass}), together with some energetic
considerations, lead to the model of a colored
black hole as a bound state of an ordinary, `bare', black hole and
a `solitonic residue', where the ADM mass of the colored black hole
of radius $R_{\Delta}$ is given by the ADM mass of the soliton plus
the horizon mass of the `bare' black hole plus the binding energy
\cite{Ashtekar:2001nx}.
Simple considerations about the behavior of the ADM masses of the
colored black holes and the solitons, together with some
expectations of this model (such as a negative binding energy)
give raise to prediction of the behavior of horizon parameters
 \cite{Ashtekar:2001nx}. Among the predictions, we have that
$\beta_{(n)}(R_{\Delta})$, as a function of
$R_\Delta$ and $n$, is a positive function,
bounded above by
$\beta_{(0)}(R_{\Delta})=1$. Besides, the curves $\beta_{(n)}(r)$,
as functions of $r$ intersect the $r=0$ axis at distinct points
between $0$ and $1$, never intersect, and have the property that
the higher $n$ is, the lower the curve. Finally, the curves, for
large value of their argument, become asymptotically tangential
to the curve  $\beta_{(0)}(R_\Delta)=1$.
One of the features of these solutions is that there is no
limit for the size of the black hole. That is, if we plot
the ADM mass of the BH as function of the radius $R_\Delta$
we get an infinite number of curves, each of them intersecting
the $R_{\Delta}=0$ line at the value of the soliton mass,
and never intersecting each other.

When one introduces extra fields, as is the case of a Higgs field in
EYM-Higgs, then there is an extra dimension-full parameter $\eta$
that brings a new mass scale into the problem. This, in turn, has as
a consequence that there is a maximum value for the horizon radius of
the black holes \cite{Nunez:1996xv,Ashtekar:2001nx}. This also indicates that
in the $M_{\rm ADM}$ vs. $R_\Delta$ plot, the curves corresponding to two
families of black holes intersect at a point. Now, recall that the
horizon Mass was obtained by integrating
$ M^{(n_o)}_{\Delta}(R_\Delta)=\frac{1}{2 G}\int_0^{R_\Delta}\beta_{(n_o)}(r)
\, \d r
$,
along the curve corresponding to the $n$th family. If now
two such curves, say the $n$-th and the $n+1$-th, intersect,
one can use this formula together with Eq.
(\ref{ymmass}), to conclude that,
\be
M^{(n+1)}_{\rm sol} - M^{(n)}_{\rm sol}=
\frac{1}{2G}\oint \beta(r)\,\d r
\label{massdiff}
\ee
where the closed counter integral is performed by first moving along the
$n$-th branch up to the crossing point and then
returning along the $n+1$-th branch back to $R_\Delta=0$
\cite{Kleihaus:2000kv,Ashtekar:2001nx}. This concludes the summary 
 of the IH formalism needed for this note.

 Now let us consider the static sector of a family of EYMH theories,
parametrized by $\eta$. As $\eta $
decreases, the second mass parameter of the theory  becomes smaller and the
upper bound on the area of
the hairy black holes increases\cite{Nunez:1996xv}. Thus we expect  that
the point of
intersection  of the $n$ and
$n+1$ branches would move towards larger  values of $R$ as $\eta$
decreases. Moreover in the limit
$\eta \to 0$ we would expect this intersection point to move towards
infinity leading to a situation
where the different branches do not intersect, as is in fact observed
to happen in pure EYM theory.
We might also argue that as  $\eta$ determines the natural scale
for the Higgs field in static
situations, we expect that as $\eta \to 0$ the static sector of EYMH theory
would have vanishing  Higgs
field and thus correspond to  the static sector  of pure EYM theory. We
will support this argument by
explicitly proving that for the case $\eta=0$ the static, purely magnetic
solutions of EYMH theory have
vanishing Higgs field. The proof is a  simple generalization of a Bekenstein
no hair theorem \cite{Bekenstein:1972hc}.

Consider an EYMH theory with scalar field $\Phi$ with potential
$V(\Phi)=\lambda (\Phi^* \Phi-\eta^2)^2$,
where $\lambda$ , and $\eta$ are constants, and $\Phi*$ is the
Hermitian conjugate of $\Phi$.
Consider a static black hole solution with time-like Killing field
$\xi^a$. The equation of motion for the scalar field is:
\be
D^aD_a \Phi -{{\partial V} \over { \partial \Phi^*}}=0
\label{eqmot}
\ee
where $D_a$ stands for the gauge and metric covariant derivative
$D_a = \nabla_a  - ie A_a^i T^i$,
with $\nabla_a$ is the metric compatible derivative operator, $A_a^i$
stand for the gauge fields,
$T^i$ for the Lie algebra generators, and $e$ is the gauge coupling constant.
For non-extremal black holes without loss of generality we can
consider that the space-time has a
bifurcate Killing Horizon with bifurcation surface $S$\cite{Racz:1995nh}. 
Let $t$  be the
Killing parameter and consider $M$
the region of space-time bounded by $\Sigma_1$ and $\Sigma_2$ hyper-surfaces
of constant $t$, $S$ and
asymptotic infinity. We multiply Eq. (\ref{eqmot}) by $\Phi^*$ and
integrate over M:
\ba
0&=&\int_M  \d^4x (\Phi^*D^aD_a \Phi -\Phi^*{{\partial V} \over { \partial
\Phi^*}})\sqrt{-g}\\ \nonumber
&=& \int_{\partial M}  \d\Sigma^a\,\Phi^*D_a\Phi -
\int_M  \d^4x \, \sqrt{-g}\,(D_a \Phi^* D_b \Phi g^{ab} \\ \nonumber
&{}& +\Phi^*{{\partial V} \over { \partial
\Phi^*}})
\ea
 Consider the case where $\eta =0$.
 The boundary integral consists of  four terms: The integrals over
$\Sigma_1$ and $\Sigma_2$ that are
equal in magnitude and opposite in sign due to the time translation
invariance and the vanishing of the integral at  infinity
because  of the fall-off conditions on $\Phi$ required from
asymptotic flatness. Finally, the
term associated with $S$ which does not contribute since the
integral  is over a lower dimensional manifold.

  Next we write the
inverse metric $g^{ab} =  -N^2
\xi^a \xi^b +h^{ab}$, where $N^{-1}$ is the norm of the Killing 
field $\xi^a$ and $h^{ab}$ is the
inverse of spatial metric on $\Sigma$. 
Using the fact that $\xi^a \nabla_a \Phi$ vanishes as long as the
solution is static, and  that $A_0^i\equiv \xi^a A_a^i =0$ corresponding to
the purely magnetic sector,
we obtain:
\be
\int_M \d^4x \,\sqrt{-g}\,
(D_a \Phi^* D_b \Phi h^{ab} +\Phi^*{{\partial V} \over { \partial
\Phi^*}})=0
\ee
The first term in the  integrand is positive semi-definite in general and
the second term becomes positive
semi-definite when $\eta= 0$. Thus in this situation  the only possibility is
$D_a\Phi =0$ and
$\Phi^*{{\partial V} \over { \partial \Phi^*}}=0 $ which for $\eta=0$
implies $\Phi=0$.  This is what we
wanted to show.

Thus, by considering the limit of the static and purely magnetic sector
of EYMH theory in the limit
$\eta \to 0$ we expect to obtain the static and purely magnetic sector of
the EYM theory. Furthermore, by
considering  Eq. (\ref{massdiff}) in this limit we obtain an equivalent
equation
(\ref{Mass}), now for the EYM case.
Let us now consider Minkowski space-time as
the zeroth  (degenerate) soliton, where the Schwarzschild branch  (the
${0}^{\rm th}$
branch)  begins, we obtain a
formula for the mass of the nth  soliton in EYM theory:
\be
 M_{\rm sol}^{(n)} =  {1\over{2G}} \oint \beta(r)\, \d r,
\ee
 where the integral is performed first along the Schwarzschild branch to
$r=\infty$ and returning to $r=0$
along the  $n^{\rm th}$ branch. This is precisely Eq. (\ref{Mass}).

\begin{table*}
\caption[]{\label{t:crit}
${\rm R}$ is the upper limit of the integration interval.
$I(n=1)$, $I(n=2)$, $I(n=2)-I(n=1)$ are the values of the integral for
one node, two nodes, and the difference of both, respectively. The complete 
output of the simulation is reported even when the numerical errors make
some of the last figures irrelevant.}
\begin{flushleft}
\begin{tabular}{c|ccc}
\hline\noalign{\smallskip}
 $\displaystyle {{\rm R} \atop \ }$  &
 $\displaystyle { I(n=1) \atop \ }$  &
 $\displaystyle { I(n=2) \atop \ }$  &
 $\displaystyle { I(n=2)-I(n=1) \atop \ }$ \\
\noalign{\smallskip}
\hline\noalign{\smallskip}
1   &  0.39103324877641  &  0.47715590602640 &  8.6122657249990D-02 \\ \hline
5   &  0.73246896963857  &  0.87112010238884 &  0.13865113275027    \\ \hline
10  &  0.78027992488956  &  0.92106063175122 &  0.14078070686166    \\ \hline
20  &  0.80423765124767  &  0.94603243120791 &  0.14179477996024    \\ \hline
30  &  0.81222747739806  &  0.95435647857942 &  0.14212900118136    \\ \hline
40  &  0.81622282131691  &  0.95851851479659 &  0.14229569347968    \\ \hline
50  &  0.81862012691719  &  0.96101573942668 &  0.14239561250949    \\ \hline
60  &  0.82021836327953  &  0.96268055681115 &  0.14246219353162    \\ \hline
70  &  0.82135997388031  &  0.96386971245266 &  0.14250973857235    \\ \hline
80  &  0.82221618794342  &  0.96476157934352 &  0.14254539140010    \\ \hline
90  &  0.82288213542783  &  0.96545525369931 &  0.14257311827148    \\ \hline
100 &  0.82341489514230  &  0.96601019325204 &  0.14259529810974    \\
\hline
110 &  0.82385079057060  &  0.96601019325204 &  0.14261344416386    \\ \hline
120 &  0.82421403734650  &  0.96684260265416 &  0.14262856530766    \\ \hline
130 &  0.82452140046910  &  0.96716276014096 &  0.14264135967186    \\ \hline
140 &  0.82478485490025  &  0.96743718084997 &  0.14265232594972    \\ \hline
150 &  0.82501318223082  &  0.96767501213306 &  0.14266182990224    \\ \hline
\noalign{\smallskip}
\hline
\end{tabular}
\end{flushleft}
\end{table*}

Now we want to establish the validity of formula (\ref{Mass}) for
EYM solitons.
For that, we show a comparison between the numerical evaluation
of the masses of the EYM solitons and the corresponding evaluation of the
left hand side of Eq. (\ref{Mass}) for the $n=1,2$ colored black holes.
To do this, we calculate numerically the functions $\beta_{(n)}(r)$ for
$n=1,2$ as a function of the radius of the horizon $R_\Delta$ by using a
fourth-order Runge-Kutta algorithm for the computation of the
configurations of black holes with $n=1,2$ nodes maintaining the truncation
error below of the accuracy $10^{-6}$ for the metric and the Yang-Mills
field of each configuration. At the same time, a fifth-order Runge-Kutta
algorithm was used for estimating the error in the values of the function
$\beta_{(n)}(r)$. Then we calculate the value of the integral (\ref{Mass})
for different values (sufficiently large) of the upper limit of
integration ${\rm R}$ and the respective accumulated error associated with
it coming of the one associated with the values of the function
$\beta_{(n)}(r)$.
The accumulated errors associated with the integral (\ref{Mass}), 
calculated as the difference between the fourth and fifth order algorithms,
for $n=1,2$ and superior limit of integration ${\rm R} = 150$, 
are $5 \times 10^{-4}$
and $9 \times 10^{-4}$ respectively (the accumulated errors
for ${\rm R} > 150$ differ of these values by a quantity of the order
$10^{-10}$
and then have practically converged to these). The black hole
radius is in the standard units given by the EYM coupling constant
(where $R=1$ corresponds to the natural length scale given by the theory).
%Finally, for $n=1,2$ and for different values of ${\rm R}$,
%we have subtracted this error of the value of the integral (\ref{Mass}).
Table \ref{t:crit} shows the values of the integral (\ref{Mass})
for $n=1,2$ and the difference between both as a
function of the upper limit of integration ${\rm R}$, using the 
fifth-order algorithm. It is important to notice from the Table
that the difference of the
integrals, which by equations (\ref{Mass}) and (\ref{massdiff}) should
approximate the difference in soliton masses, has indeed a faster 
convergence than the integrals by themselves. We have restricted the
upper limit in the integral to ${\rm R = 150}$ given that the difference in
masses had practically converged to the reported value.

From the values given in Table~\ref{t:crit},
it is clear that, For ${\rm R}$ sufficiently large,,
the integrals for $n=1,2$ (and the difference of both) are
approaching to the corresponding soliton masses, $0.82864698216$ and
$0.97134549426$ for $n=1,2$ respectively (and the difference of them, 
$0.142698512$), as reported in \cite{Breitenlohner:1994es}. 

To conclude, within the uncertainty of our numerical 
calculation, the soliton mass formula (\ref{Mass}) provides 
the ADM mass of the EYM system.

%\section*{acknowledgments}

We would like to thank A. Ashtekar for a careful reading of the
manuscript and J. Smoller for pointing out some references.
The authors were supported in part by DGAPA-UNAM grant
IN121298 and by CONACyT Proy. Ref. J32754-E and 32272-E.
U.N. would like to thank CONACyT for finantial support.

\end{document}